\documentclass[aps, prl, reprint, letterpaper, longbibliography, superscriptaddress]{revtex4-1}

\usepackage{amsmath,amssymb,amsfonts,latexsym,color,dcolumn,bm,mathtools,amsbsy}
\usepackage[english]{babel}
\usepackage{graphicx}
\usepackage[utf8]{inputenc}
\usepackage{textcomp}
\usepackage{siunitx}
\usepackage[colorlinks,urlcolor=blue,citecolor=blue,unicode]{hyperref}
\hypersetup{
	pdftitle={Cavity quantum electrodynamics with frequency-dependent reflectors},
	pdfauthor={O. \v{C}ernot\'ik, A. Dantan, C. Genes}
}
\usepackage{verbatim}

\newcommand{\nbar}{\ensuremath{\bar{n}}}
\newcommand{\Hint}{\ensuremath{H_\mathrm{int}}}
\newcommand{\Hc}{\ensuremath{\mathrm{H.c.}}}
\newcommand{\diag}{\ensuremath{\mathrm{diag}}}

\newcommand{\inpt}{\ensuremath{\mathrm{in}}}
\newcommand{\out}{\ensuremath{\mathrm{out}}}
\newcommand{\eff}{\ensuremath{\mathrm{eff}}}
\newcommand{\avg}[1]{\langle #1\rangle}

\newcommand{\opt}{\ensuremath{\mathrm{opt}}}

\begin{document}


\title{Cavity Quantum Electrodynamics with Frequency-Dependent Reflectors}

\author{Ond\v{r}ej \v{C}ernot\'ik}
\altaffiliation{Present address: Department of Optics, Palack\'y University, 17. listopadu 12, 771 46 Olomouc, Czech Republic}
\email{ondrej.cernotik@upol.cz}
\affiliation{Max Planck Institute for the Science of Light, Staudtstra\ss{}e 2, 91058 Erlangen, Germany}

\author{Aur\'elien Dantan}
\affiliation{Department of Physics and Astronomy, University of Aarhus, DK-8000 Aarhus C, Denmark}

\author{Claudiu Genes}
\affiliation{Max Planck Institute for the Science of Light, Staudtstra\ss{}e 2, 91058 Erlangen, Germany}

\begin{abstract}
    We present a general framework for cavity quantum electrodynamics with strongly frequency-dependent mirrors. The method is applicable to a variety of reflectors exhibiting sharp internal resonances as can be realized, for example, with photonic-crystal mirrors or with two-dimensional atomic arrays around subradiant points. Our approach is based on a modification of the standard input--output formalism to explicitly include the dynamics of the mirror's internal resonance. We show how to directly extract the interaction tuning parameters from the comparison with classical transfer matrix theory and how to treat the non-Markovian dynamics of the cavity field mode introduced by the mirror's internal resonance. As an application within optomechanics, we illustrate how a non-Markovian Fano cavity possessing a flexible photonic-crystal mirror can provide both sideband resolution as well as strong heating suppression in optomechanical cooling. This approach, amenable to a wide range of systems, opens up possibilities for using hybrid frequency-dependent reflectors in cavity quantum electrodynamics for engineering novel forms of light–matter interactions.
\end{abstract}

\date{\today}

\maketitle



\begin{figure}[t]
    \centering
    \includegraphics[width=0.95\linewidth]{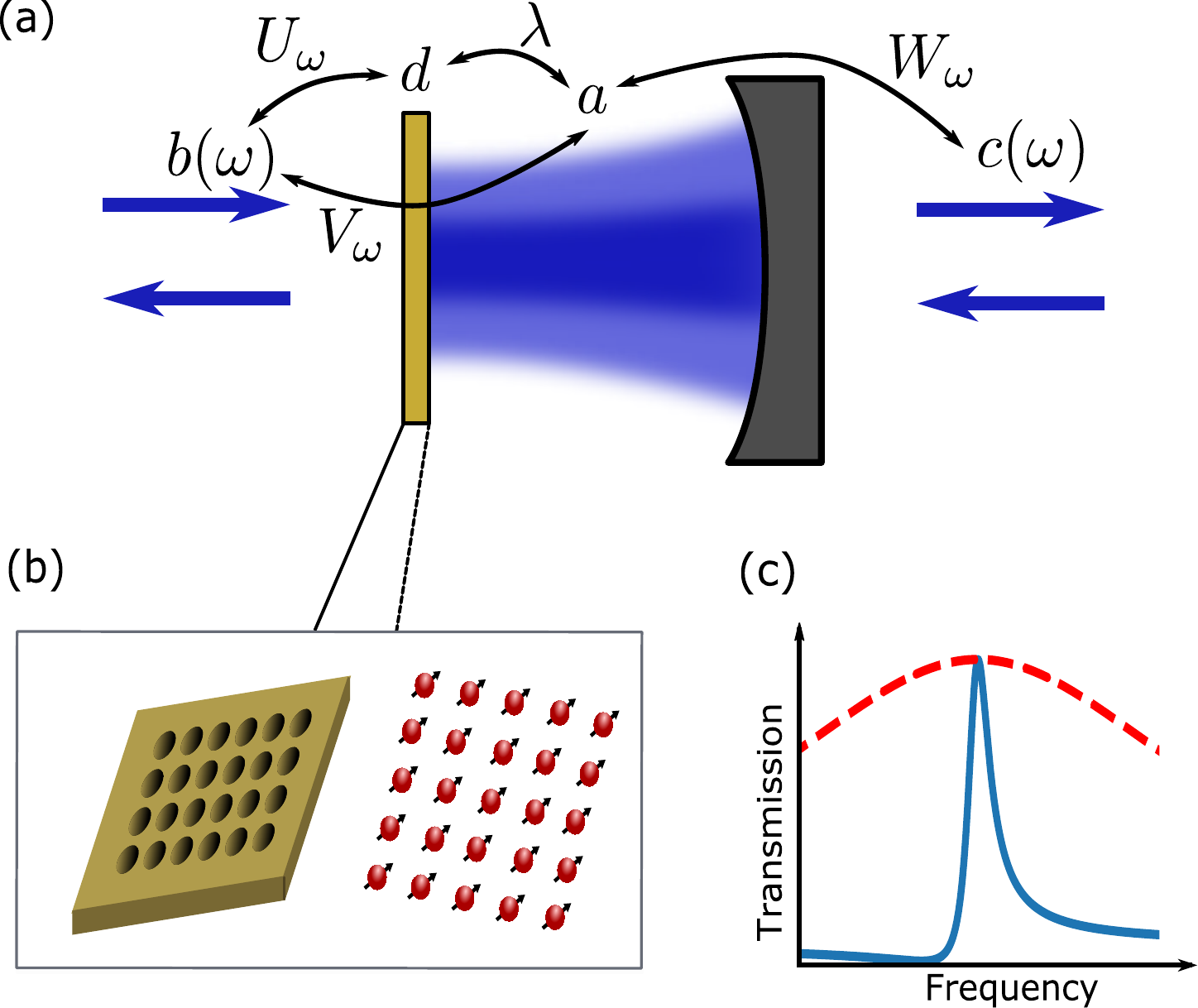}
    \caption{\label{fig:Scheme}\textit{Non-Markovian cavity}
        (a) The cavity mode $a$ interacts with two external continua $b(\omega)$, $c(\omega)$ and an internal mirror mode $d$.
        (b) Possible realizations of the resonant mirror: dielectric membrane patterned with a photonic crystal structure, two-dimensional array of atoms.
        (c) Transmission of a cavity with a resonant end-mirror (solid blue line) exhibiting an asymmetric Fano profile and with linewidth significantly reduced compared to a Markovian cavity with conventional mirrors (dashed red line).}
\end{figure}

A standard platform for cavity quantum electrodynamics (CQED)~\cite{Berman1994,Walther2006,Haroche2013} is the linear Fabry--P\'erot resonator;
one generally assumes two macroscopic, highly reflecting mirrors that define spatially-localized frequency-resolved resonances inside the cavity.
A full quantum description of the cavity mode dynamics can be derived in the form of a Langevin equation $\dot{a}(t) = -i\omega_a a(t)-\kappa a(t)+\sqrt{2\kappa} a_\inpt(t)$
where $a$ is the annihilation operator of the field mode with frequency $\omega_a$ and decay rate $\kappa$ and $a_\inpt(t)$ describes delta-correlated input noise encompassing the effect of the coupling to the continuum of outside modes~\cite{Gardiner2004}. The solution, combined with the input--output relation $a_\out(t)=a_\inpt(t) -\sqrt{2\kappa} a(t)$, describes the quantum properties of the continuous outgoing light field $a_\out(t)$.
In such a case, the quantum dynamics of the cavity field are Markovian, the coupling to the continuum of outside modes giving rise to an exponential time decay of the intracavity field.
A critical step in this derivation lies in assuming that the reflectivity of the mirrors is essentially flat around the resonance frequency of interest.\\
\indent Many scenarios, however, strongly depart from this situation as end-mirrors can be made of reflective materials exhibiting enhanced linear or nonlinear response around frequencies corresponding to sharp internal modes (Fig.~\ref{fig:Scheme}).
In two-dimensional systems, these effects can be achieved by patterning a subwavelength grating or a photonic-crystal structure onto a dielectric membrane~\cite{Miroshnichenko2010,Chang-Hasnain2012,Zhou2014,Limonov2017};
other systems can be formed by semiconducting monolayers~\cite{Zeytinoglu2017,Back2018,Scuri2018} or two-dimensional arrays of atoms trapped in optical lattices~\cite{Bettles2016,Shahmoon2017,Wild2018}.
Using such metamaterials with a strongly frequency-dependent response as end-mirrors in Fabry--P\'erot resonators has been shown to result in asymmetric transmission profiles potentially much narrower than those obtained with frequency-independent mirrors of comparable reflectivity~\cite{Naesby2018} (Fig.~\ref{fig:Scheme}(c)).\\
\indent This manuscript addresses provides a generalized approach to CQED with mirrors possessing sharp internal resonances.
Our strategy is based on an extension of the standard derivation of cavity input--output relations that includes, as an additional degree of freedom, the quantum dynamics of the mirror's internal mode and its coupling to both the cavity field and the continuum outside (Fig.~\ref{fig:Scheme}).
The result is a compact Langevin equation for the cavity field,
\begin{equation}\label{eq:NonMarkovian}
    \dot{a}(t) = -i\omega_a a(t)-(\kappa_\eff\ast a)(t)+(\kappa_\inpt \ast a_\inpt)(t),
\end{equation}
with a generalized decay rate and input noise encompassing [via the convolution $(f\ast g)(t)$] non-Markovian effects stemming from the delayed response of the field to the input noise. The explicit dependence of $\kappa_\eff(t)$ and $\kappa_\inpt(t)$ on the system parameters is obtained by identifying the predicted transmission of the quantum model to classical results obtained via a transfer matrix approach (simulation of Maxwell's equations in one dimension). We illustrate this procedure on the particular example of a cavity with a photonic-crystal membrane (Fano mirror) by using a generic frequency-dependent polarizability derived for this system by Fan {\it et al.}~\cite{Fan2003}. We then use this quantum coupled modes approach to illustrate a non-Markovian regime for cavity optomechanics (OM) where the mirror-induced reduction in the cavity linewidth combined with the suppression of the heating sideband by the asymmetric cavity profile can provide enhanced cooling dynamics.
\paragraph{\textbf{Non-Markovian Langevin equations}.}
A standard quantum optics approach~\cite{Gardiner2004} to deriving Langevin equation of motion for a mode $a$ of an optical cavity with leaky mirrors starts from the complete Hamiltonian $H = H_S + H_B + \Hint$ of the system, bath and their interaction. The free evolution is given by $H_S = \omega_a a^\dagger a$ (for system), $H_B = \int d\omega\,\omega[b^\dagger(\omega)b(\omega)+c^\dagger(\omega)c(\omega)]$ (accounting for the external free fields $b(\omega)$, $c(\omega)$) while the mutual coupling is $\Hint = \int d\omega\,i[V_\omega b(\omega)+W_\omega c(\omega)]a^\dagger + \Hc$.  The model assumes an exchange interaction with frequency-dependent rates $V_\omega$ and $W_\omega$ describing tunnelling of photons between the cavity and the free fields through the end-mirrors. Formal integration of the evolution of the outside modes leads to an effective dissipative dynamics for the cavity mode $a$. A critical step is the simplifying assumption that $V_\omega$ and $W_\omega$ are weakly dependent on frequency such that only their value at the cavity resonance is relevant.\\ For the situation depicted in Fig.~\ref{fig:Scheme}(a) the right-hand mirror is frequency-independent, ensuring that $W_\omega$ is practically constant around the resonance. The same is however not true for the sharply peaked frequency response of the left-hand mirror. Therefore, the coupling between $a$ and the mirror-filtered outside modes $V_\omega$ is necessarily strongly frequency-dependent. The elimination of the outside modes strictly following the standard formalism is therefore not straightforward. However, the approach of Ref.~\cite{Fan2003} suggests that the task can be greatly simplified by explicitly including the time dynamics of the left-hand mirror's internal mode.
This modification is done by adding a quantized mode $d$ directly coupled to the cavity and to the outside continuum.
We adjust the free Hamiltonian to include the free evolution $\omega_d d^\dagger d$ of mode $d$ at frequency $\omega_d$ and its coupling to the cavity mode $\lambda(a^\dagger de^{i\phi}+d^\dagger ae^{-i\phi})$ with complex strength~$\lambda e^{i\phi}$. The interaction Hamiltonian is also modified by a term $U_\omega b(\omega)d^\dagger + \Hc,$ in order to provide a mechanism for direct exchange of photons between the mirror mode and the left-side modes $b(\omega)$ at rate $U_\omega$. As all bosonic modes are now directly coupled to their baths, it is now allowed to consider the tunnelling rates $U,V,W\in\mathbb{R}$ to be frequency independent (hence, we drop the subscript $\omega$). Tunneling will then give rise to loss rates $\kappa_L = \pi V^2$, $\kappa_R = \pi W^2$, $\gamma = \pi U^2$.
The dynamics are then described by the Langevin equations (see Supplemental Material-SM~\cite{Supplement})
\begin{subequations}\label{eq:EOM}
\begin{align}\label{eq:EOMCavity}
    \dot{a} &= -(i\omega_a+\kappa)a - \mathcal{G}_+d + \sqrt{2\kappa_L}b_\inpt + \sqrt{2\kappa_R}c_\inpt, \\
    \dot{d} &= -(i\omega_d+\gamma)d - \mathcal{G}_-a + \sqrt{2\gamma}b_\inpt\label{eq:EOMFano},
\end{align}
\end{subequations}
where we introduced $ \mathcal{G}_\pm = i\lambda e^{\pm i\phi}+\sqrt{\kappa_L\gamma}$, the total decay rate $\kappa = \kappa_L+\kappa_R$ and the input fields on the left and right denoted by $b_\inpt$ and $c_\inpt$.
The input fields have the usual non-vanishing correlation functions $\avg{b_\inpt(t)b_\inpt^\dagger(t')} = \delta(t-t')$ (and similarly for $c_\inpt(t)$), while all other correlations vanish.
The associated output fields follow the input--output relations $b_\out = b_\inpt-\sqrt{2\kappa_L}a-\sqrt{2\gamma}d,$ and $c_\out = c_\inpt -\sqrt{2\kappa_R}a$. As we are interested in the dynamics of the cavity mode alone, we can formally integrate the evolution of the mirror mode
and replace it back in Eq.~\eqref{eq:EOMCavity}.
We thus recover the non-Markovian cavity dynamics of Eq.~\eqref{eq:NonMarkovian} where we can identify the memory kernels
\begin{subequations}\label{eq:Coefficients}
\begin{align}\label{eq:kappaeff}
    \kappa_\eff(t) &= 2\kappa_L\delta(t)-\mathcal{G}_+\mathcal{G}_-e^{-(i\omega_d+\gamma)t},\\
    \kappa_\inpt(t) &= 2\sqrt{2\kappa_L}\delta(t)-\mathcal{G}_+\sqrt{2\gamma}e^{-(i\omega_d+\gamma)t}.
\end{align}
\end{subequations}
Notice that Eq.~\eqref{eq:NonMarkovian} is written for a single input noise. The double-sided cavity considered here exhibits a Markovian response to $c_\inpt $ and a non-Markovian response to $b_\inpt$ characterized by the effective time-dependent decay rate $\kappa_\eff(t)$ and associated response with  $\kappa_\inpt(t)$ to the left-side input noise field $b_\inpt$.
The outgoing field $b_\out$ will also show a memory effect as it includes a time-delayed response via the mirror mode of both $a(t)$ and $b_\inpt(t)$ according to
\begin{subequations}\label{eq:InOutNonMark}
\begin{align}
    b_\out &= (\kappa_\out\ast b_\inpt)(t) - (\kappa_\inpt'\ast a)(t),\\
    \kappa_\out(t) &= 2\delta(t)-2\gamma e^{-(i\omega_d+\gamma)t},\\
    \kappa_\inpt'(t) &= 2\sqrt{2\kappa_L}\delta(t)-\mathcal{G}_-\sqrt{2\gamma}e^{-(i\omega_d+\gamma)t}.
\end{align}
\end{subequations}
The expressions above are completely general as they do not involve any assumptions on the particular nature of the mirror's internal mode. The parameters $\lambda,\kappa_L,\kappa_R,\omega_a$ are to be determined from comparisons of the predictions of the coupled mode quantum model for cavity transmission, reflection with classical transfer matrix methods. In general, $\kappa_L$ is not the dissipation rate of the cavity quasi-mode but rather a very large rate of coupling to the vacuum modes; also, the cavity quasi-mode frequency will be defined by the resonance of the mirror. Furthermore, against intuition, the limit of frequency-insensitive mirrors, reached for infinite $\gamma$ leads to infinite couplings $\mathcal{G}_{\pm}$. However, as we will illustrate with photonic crystal mirrors, terms such as $\gamma/2 e^{-\gamma |t|}$ tend to $\delta(t)$ for infinite $\gamma$, thus reproducing the expected instantaneous $\kappa_\eff(t)\propto\delta(t)$ response characteristic of Markovian dynamics.
\paragraph{\textbf{Transfer matrix comparison}.}
We now describe a roadmap to extract the parameters of the coupled-mode model by matching its predictions to a classical transfer matrix calculation. The transmission coefficient $t(\omega)=\avg{c_\out(\omega)}/\avg{b_\inpt(\omega)}=\sqrt{\kappa_R} \avg{a(\omega)}/\avg{b_\inpt(\omega)}$ can be analytically obtained by solving the equations of motion~\eqref{eq:EOM} in the frequency domain such that
\begin{widetext}
\begin{equation}
    t(\omega) = \frac{2i\sqrt{\kappa_R}[e^{i\phi}\lambda\sqrt{\gamma}-(\omega_d-\omega)\sqrt{\kappa_L}]}{\lambda^2+\kappa_R\gamma-(\omega_d-\omega)(\omega_a-\omega)+i[\gamma(\omega_a-\omega)+\kappa(\omega_d-\omega)-2\lambda\sqrt{\kappa_L\gamma}\cos\phi]}.
\end{equation}
\end{widetext}
The transfer matrix approach, on the other hand, consists in solving the classical one-dimensional wave propagation equation in a one-dimensional setup with two mirrors parametrized by polarizabilities $\zeta_R$ and $\zeta_L(\omega)$.
In linear response theory, one can find the transmission function of the setup for any incoming plane wave at a given frequency $\omega$.
As detailed in SM~\cite{Supplement}, we get the classical transmission coefficient
\begin{equation}\label{eq:TransTM}
    \tilde{t}(\omega)
    = \frac{1}{(1-i\zeta_R)[1-i\zeta_L(\omega)]e^{-i\theta}+\zeta_R\zeta_L(\omega)e^{i\theta}};
\end{equation}
here, $\theta = \frac{1}{2}[(\omega_d-\omega)/\Gamma + \arctan(1/\zeta_R)+\arctan(1/\zeta_0)]$ with $\Gamma = c/2L$ denoting the free spectral range and $\zeta_0 = \zeta_L(\omega = \omega_d)$.
Comparison of the two expressions, $t(\omega)$ and $\tilde{t}(\omega)$, gives the free parameters of the coupled mode description. We will exemplify this procedure below in the particular case of a mirror formed by a photonic-crystal membrane.

\begin{figure}[b]
    \centering
    \includegraphics[width=1\linewidth]{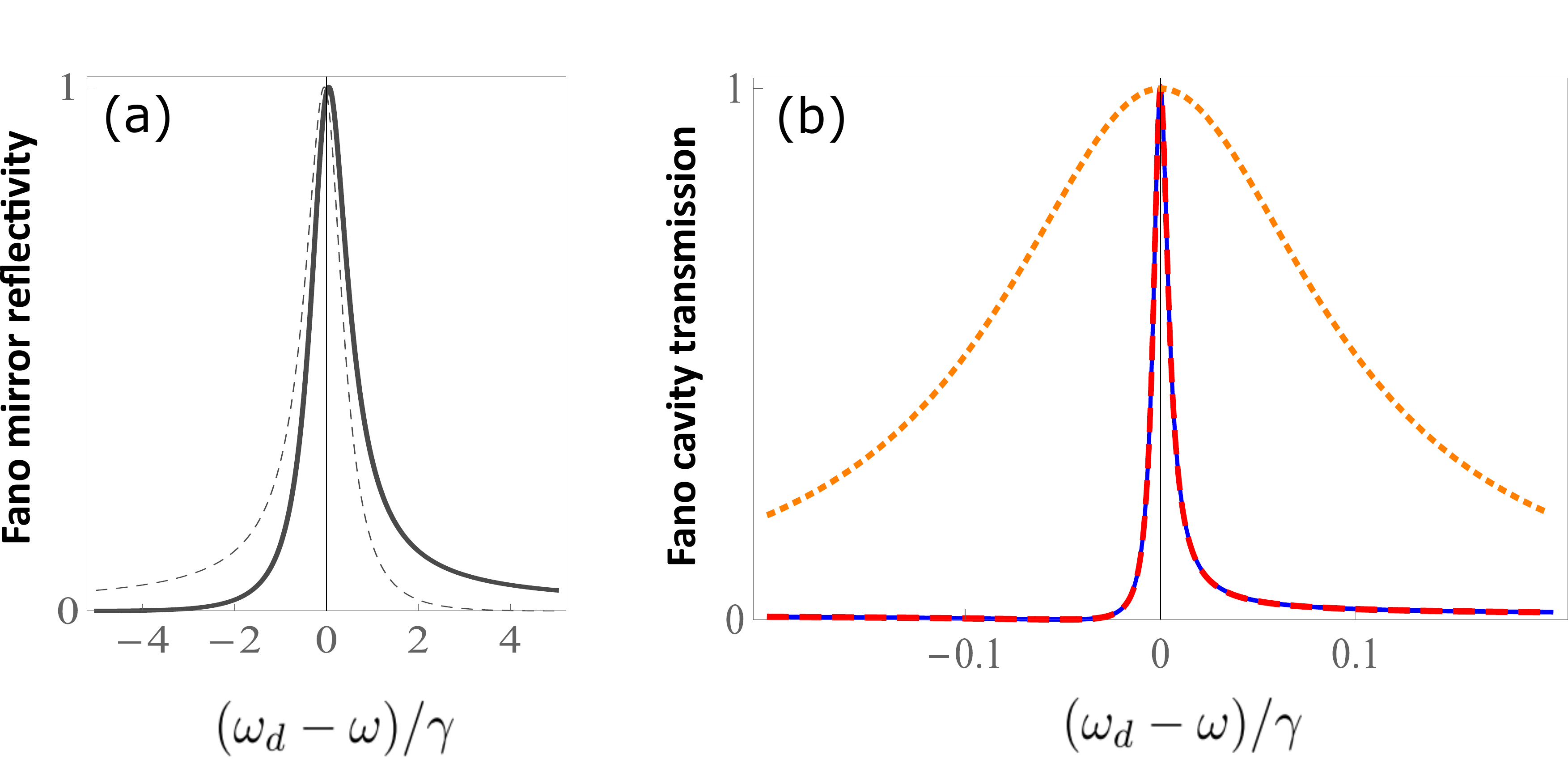}
    \caption{\label{fig:Comparison}
        \textit{Non-Markovian Fano cavities}.
        (a) Fano mirror reflectivity $|i\zeta_L(\omega)/[1-i\zeta_L(\omega)]|^2$ with $\zeta_0 = 10$ for $s = -1$ (solid line) and $s = 1$ (dashed line).
        (b) The dashed orange line shows a broad Markovian cavity transmission for $\zeta_0 = \zeta_R = 10$ and $\gamma\to\infty$. In contrast, coupled-mode theory (solid blue line) and transfer matrix calculations (dashed red line) for $\gamma = \Gamma/10$ and $s =-1$ show a considerable linewidth reduction accompanied with an asymmetric Fano profile.}
\end{figure}

\paragraph{\textbf{Photonic crystal mirrors}.}
Fano resonances in thin dielectric membranes patterned with subwavelength photonic-crystal structures may generically be described by the polarizability~\cite{Naesby2018,Fan2003}
\begin{equation}\label{eq:zetaL}
    \zeta_L(\omega) = \zeta_0\frac{\gamma-2s(\omega_d-\omega)/\zeta_0}{\gamma+2s\zeta_0(\omega_d-\omega)},
\end{equation}
with $s$ specifying the orientation of the Fano resonance [see Fig.~\ref{fig:Comparison}(a)].
We assume $\zeta_0\gg 1$, leading to a small off-resonant value $\zeta_L\to 1/\zeta_0$ at $\omega\to\pm\infty$.
The left mirror is, therefore, a poor reflector of its own and only the presence of its internal mode gives rise to a large effective reflectivity on resonance.
To extract the parameters of the coupled mode quantum theory from the classical transfer matrix results we follow three steps: i) identifying the zero of the cavity transmission , ii) taking the limit $\gamma\gg\Gamma$ (Markovian cavity) and iii) taking the limit of fully non-Markovian behavior with $\gamma\ll\Gamma$. The three conditions lead to~\cite{Supplement}
\begin{equation}\label{eq:FanoZero}
    \lambda = -s\sqrt{\kappa_0\gamma},\qquad
   \omega_a = \omega_d -2s\sqrt{\kappa_0\kappa_L}.
\end{equation}
The decay rates can be identified as  $\kappa_0 = \Gamma/2\zeta_0^2$, $\kappa_L = 2\Gamma$, and $\kappa_R = \Gamma/2\zeta_R^2$ and $\phi = 0$. Notice that $\kappa_0$ and $\kappa_R$ are the expected loss rates of photons to the continuum for frequency-independent mirrors with fixed $\zeta_0,\zeta_R$. The quantity $\kappa_L$ instead describes the high loss of the cavity to the outside modes: far from the mirror resonance the effective finesse is proportional to $\zeta_R\zeta_L(\omega\to\pm\infty)=\zeta_R/\zeta_0$, which is of order unity and describes a quasi-mode with linewidth of the order of the free spectral range.

In the limit $\gamma\gg\Gamma$ Markovian dynamics is expected. With $\mathcal{G}_{\pm}=\mathcal{G}=-is\sqrt{\kappa_0\gamma}+\sqrt{\kappa_L\gamma}$ one can now compute the terms appearing in Eq.~\eqref{eq:kappaeff} and take the limit $\gamma\to\infty$~\cite{Supplement}. We obtain $\kappa_\eff(t) = 2\kappa_0\delta(t)-4is\sqrt{\kappa_0\kappa_L}\delta(t)$, which, after time integration, leads to the expected result that the mode $a$ is shifted by the quantity $2s\sqrt{\kappa_0\kappa_L}$ to fit the mirror frequency $\omega_d$ and effectively decays at rate $\kappa_0$ through the left mirror (in addition to the decay at rate $\kappa_R$ through the frequency independent right-mirror). The expected Lorentzian spectrum with width $\kappa_0+\kappa_R$ is then observed [Fig.~\ref{fig:Comparison}(b)].

In the limit $\gamma\ll\Gamma$ the Fano mirror leads a narrower cavity transmission profile with effective linewidth $\kappa_\eff = \gamma/\zeta_0^2$ much smaller than the linewidth $\Gamma/\zeta_0^2$ of a standard cavity with fixed $\zeta_0$. Moreover, the Fano interference manifests itself in a non-Markovian asymmetry with a zero in the transmission, tunable in position by the design of the photonic crystal ($s = \pm 1$). Both effects are illustrated in Fig.~\ref{fig:Comparison}(b). Perfect agreement between exact transfer matrix simulations and the coupled mode theory is obtained expect in an intermediate regime $\gamma\sim\Gamma$ where the mirror's reflectivity spectrum becomes broad enough to interact with multiple cavity modes~\cite{Supplement}. However, since the transmission profile becomes much broader we do not focus on this regime.
\begin{figure}[t]
    \centering
    \includegraphics[width=0.92\linewidth]{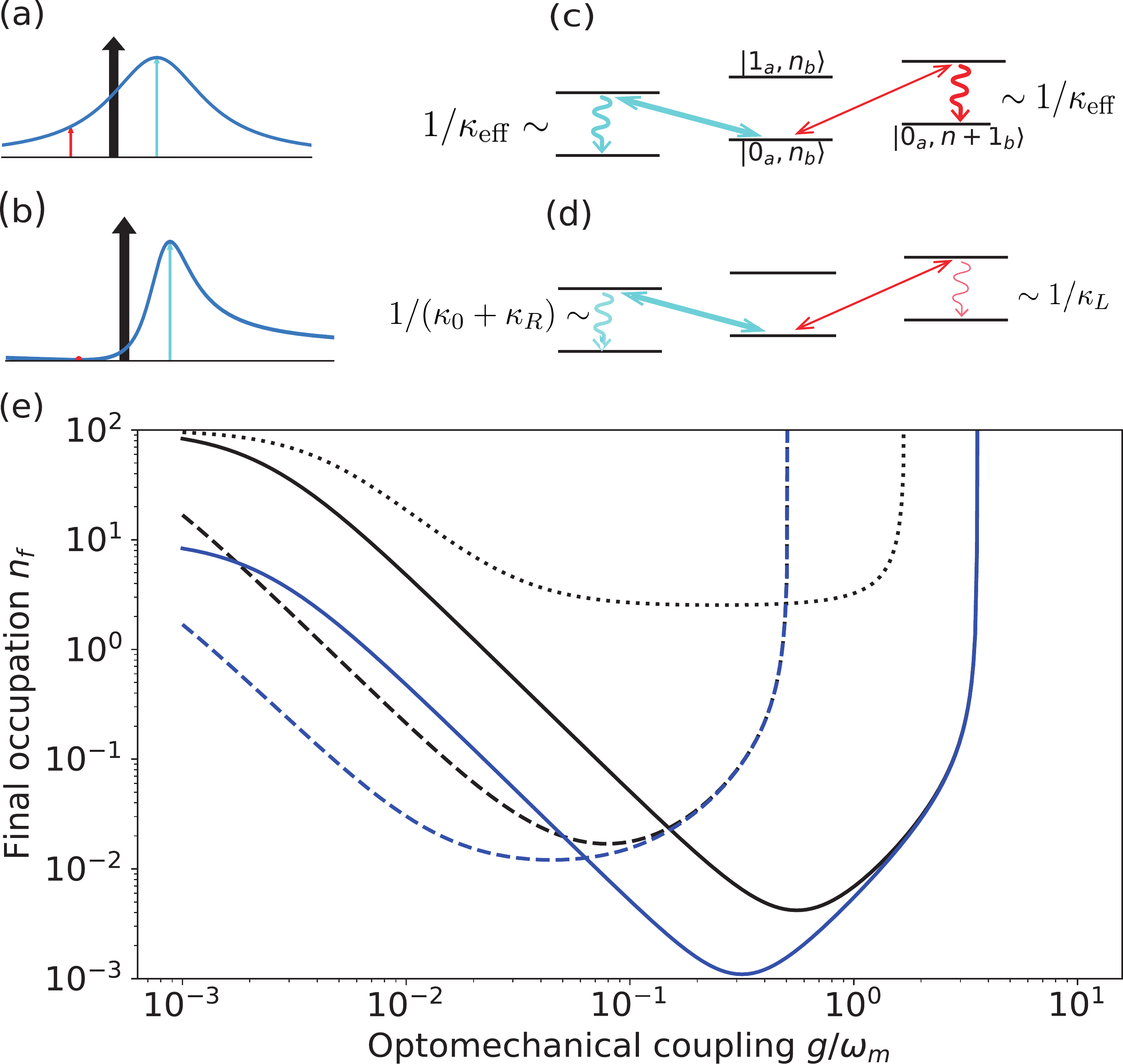}
    \caption{\label{fig:Cooling}
        \textit{OM cooling}.
        (a) Sideband cooling for Markovian cavities.
        Laser drive red-detuned from the resonance (thick black arrow) leads to enhancement of the cooling sideband (cyan) and minimization of the heating processes (red).
        (b) Non-Markovian cavity cooling showing an almost complete inhibition of Stokes scattering.
        (c,d) Level scheme for cooling with (c) Markovian and (d) non-Markovian cavities.
        (e) Final mechanical occupation $n_f$ with increasing $g$.
        The dotted line shows unresolved sideband cooling for a cavity with frequency independent mirror with $\zeta_0 = 10$. In comparison the dashed lines show the performance of a Markovian cavity with linewidth $\kappa_\eff = \gamma/2\zeta_0^2$ for $\nbar = 10^2$ (black) and $\nbar = 10$ (blue). Finally, the non-Markovian cavity performance is shown by the solid lines for $\nbar = 10^2$ (black) and $\nbar = 10$ (blue). Other parameters are $\Gamma = 1000\omega_m$, $\zeta_0 = 10$, $\gamma = 40\omega_m$, and $Q_m = \omega_m/\gamma_m = 10^6$.
    }
\end{figure}
\paragraph{\textbf{Optomechanical sideband cooling}.}
Let us analyze the performance of a cavity OM setup with frequency-dependent mirrors in the non-Markovian regime with $\gamma\ll\Gamma$. Conventional OM employs a radiation pressure Hamiltonian, $H_\mathrm{RP} = g_0 a^\dagger aq$ (where $q$ is the dimensionless position quadrature of the vibrating mirror), which, in the standard weak phonon-photon coupling, can be transformed into a linearized interaction $H_\mathrm{lin} = g(a+a^\dagger)q$~\cite{Aspelmeyer2014}.
The coupling rate $g = g_0\alpha$ is enhanced by the field amplitude in steady state $\alpha\gg 1$.
A decomposition into creation and annihilation operators, $q = (b+b^\dagger)/\sqrt{2}$, shows two contributions: i) a beam-splitter interaction $g(a^\dagger b + b^\dagger a)$ responsible for state transfer or cooling and ii) a two-mode squeezing interaction $g(ab+a^\dagger b^\dagger)$ resulting in entanglement or heating.
When the frequency of the driving laser $\omega_L$ fulfills $\omega_L=\omega_a- \omega_m$ [see Fig.~\ref{fig:Cooling}(a,c)], the beam-splitter interaction is resonant and the cavity field serves as an additional zero-temperature bath for the mechanical resonator; phonons are converted into photons which leave the cavity through its end-mirrors.
The heating contribution sets a lower limit to achievable final occupancies which scales with the sideband ratio $\kappa/\omega_m$, such that ground state cooling is more efficient in the resolved sideband regime, $\kappa<\omega_m$~\cite{Marquardt2007,Wilson-Rae2007}.

An inspection of the cavity transmission profile in Fig.~\ref{fig:Cooling}(b) reveals two advantages of using frequency-dependent reflectors: i) the reduced linewidth enables a better sideband resolution and ii) the interference between the mirror and cavity modes leads to vanishing density of modes in a small frequency window that can result in the suppression of Stokes scattering. To assess these expectations, we derive perturbative expressions for the cooling rate and final occupancy starting from the Langevin equations for the mechanical resonator,
\begin{equation}
    \dot{q} = \omega_m p,\qquad
    \dot{p} = -\omega_m q - \gamma_m p + f_\mathrm{th} - F,
\end{equation}
which describe its free evolution at frequency $\omega_m$ and damping at a rate $\gamma_m$;
$f_\mathrm{th}$, with the correlation function $\avg{f_\mathrm{th}(t)f_\mathrm{th}(t')} = \gamma_m(2\nbar+1)\delta(t-t')$, is the associated thermal noise operator.

Cooling from an initial high temperature (characterized by the average thermal occupation $\nbar\gg 1$) is achieved by the Langevin radiation pressure force $F = g(a+a^\dagger)$; its effect can be quantified using the cooling rate $\Gamma_\mathrm{cool} = \frac{1}{2}[S_F(\omega_m)-S_F(-\omega_m)]$
with $S_F(\omega)$ denoting the spectral density of $F$ at frequency $\omega$. We find~\cite{Supplement}
\begin{widetext}
\begin{equation}
    S_F(\omega) = \frac{2g^2\kappa_R[\gamma^2+(\omega-\Delta_d)^2]+2g^2[\gamma\sqrt{\kappa_0}-s\sqrt{\kappa_L}(\omega-\Delta_d)]^2}{[(\kappa+\gamma)(\omega-\Delta_d)]^2 + [\gamma(\kappa_0+\kappa_R)-(\omega-\Delta_d)(\omega-\Delta_d+2s\sqrt{\kappa_0\kappa_L})]^2}.
\end{equation}
\end{widetext}
Heating is minimized for $ \Delta_d = \omega_d-\omega_L = -\omega_m-s\gamma\sqrt{\kappa_0/\kappa_L} = -\omega_m -s\gamma/2\zeta_0$; its rate is $S_F(-\omega_m)=2g^2 \kappa_R\zeta_0^2/\Gamma^2$ and vanishes for a one-sided cavity (with $\kappa_R = 0$).
Cooling is maximized for the detuning $\Delta_d = \omega_m$ at the rate $S_F(\omega_m) = 2g^2/(\kappa_0+\kappa_R)$.
Ideally, we will both minimize the Stokes and maximize the anti-Stokes scattering;
this can be achieved for $s = -1$ and optimal mirror linewidth $\gamma_\opt = 4\zeta_0\omega_m$, as illustrated in Fig.~\ref{fig:Cooling}(b).
Note that one can, similarly, suppress the anti-Stokes scattering for $\Delta_d = -\omega_m$, with the interaction phase $s = +1$. Such a setting might offer an advantage when strong amplification of the mechanical motion is necessary, for example for reaching mechanical limit cycles~\cite{Lorch2014}. The comparison with exact numerical simulations~\cite{Supplement} is presented in Fig.~\ref{fig:Cooling}(c). First, we plot the final occupancy for standard cooling in a Markovian cavity with the left mirror characterized by a fixed $\zeta_0$ (dotted line). Then, we consider a non-Markovian cavity with a Fano mirror having $\gamma\ll\Gamma$ and observe that the cooling performance is greatly enhanced by a combination of linewidth narrowing and non-Markovian cancellation of heating rates (full lines). The Fano effect of Stokes suppression is evident when comparing with a Markovian cavity with the same narrow linewidth $\kappa_\eff = \gamma/2\zeta_0^2$.
The non-Markovian cavity exhibits a smaller cooling rate $\propto1/(\kappa_0+\kappa_R)$ than the equivalent Markovian one $\propto1/\kappa_\eff$. The advantage lies in the much lower final occupancies achievable owing to the suppression of heating.
\paragraph{\textbf{Discussion}.}
Besides sideband cooling the reduced Stokes scattering obtained with frequency-dependent reflectors can also improve the fidelity of state transfer~\cite{Palomaki2013,Weaver2017} and frequency conversion in optomechanical systems~\cite{Midolo2018}.
Suppression of anti-Stokes scattering can be beneficial for more efficient amplification of electromagnetic fields or mechanical motion~\cite{Ockeloen-Korppi2016,Toth2017} and for reaching mechanical limit cycles~\cite{Lorch2014}.
The reduced cavity linewidth can enable all these applications also in micromechanical cavities~\cite{Flowers2012,Shkarin2014,Naesby2018}, which are typically precluded from reaching the resolved sideband regime owing to their large linewidths. Photonic crystal membranes received attention lately as a possible platform for quantum OM~\cite{Bui2012,Kemiktarak2012,Stambaugh2015,Bernard2016,Norte2016,Chen2017a,Nair2018,Gartner2018};
we show how the effects that may arise owing to their frequency-dependent reflectivity can be rigorously described quantum mechanically. Our results can furthermore be applied to other platforms as well, such as arrays of trapped atoms~\cite{Bettles2016,Shahmoon2017,Shahmoon2018,Shahmoon2018a} or semiconductor membranes~\cite{Back2018,Scuri2018,Zhou2019}.

In the future, it would be interesting to analyze how cavity QED with quantum emitters is modified by using frequency-dependent end-mirrors.
Here, the modified width and shape of the cavity mode is expected to lead to nontrivial dynamics of strongly coupled polaritons or modified photon statistics. Coupling ensembles of quantum emitters to non-Markovian cavities could as well lead to modified super- or subradiance~\cite{Plankensteiner2017} and lasing~\cite{Mork2014,Yu2017}. Frequency-dependent reflection has also been employed in photonic crystals with quantum dots~\cite{Yu2015,Yu2017}
and it can be used in one-dimensional systems with emitters in optical waveguides~\cite{Chang2012,Goban2012} and in superconducting~\cite{Lalumiere2013,VanLoo2013} or plasmonic systems~\cite{Chang2006,Akimov2007}.


\begin{acknowledgements}
    We gratefully acknowledge financial support from the Max Planck Society and the Velux Foundations. We thank Christian Sommer for relevant comments on the manuscript.
\end{acknowledgements}

\bibliography{ResonantOptomechanics}


\clearpage
\setcounter{equation}{0}
\renewcommand \theequation {S\arabic{equation}}
\setcounter{figure}{0}
\renewcommand \thefigure {S\arabic{figure}}

\begin{widetext}
\begin{center}
    \textbf{Supplemental Material: Cavity Quantum Electrodynamics with Frequency-Dependent Reflectors}
\end{center}

\section{Coupled-mode theory for cavities with frequency-dependent reflectors}

Here, we present a derivation of the Langevin equations and input--output relations for the coupled mode description of a cavity with a resonant end mirror.
We start from the full Hamiltonian describing the dynamics of the cavity field, the mirror mode, and the two continua to the left and right of the cavity $H = H_S+H_B+\Hint$ as presented in the main text:
\begin{subequations}
\begin{align}
    H_S &= \omega_a a^\dagger a+\omega_d d^\dagger d + \lambda(a^\dagger de^{i\phi}+d^\dagger ae^{-i\phi}),\\
    H_B &= \int d\omega\,\omega[b^\dagger(\omega)b(\omega)+c^\dagger(\omega)c(\omega)],\\
    \Hint &= \int d\omega\,i[V_\omega b(\omega)+W_\omega c(\omega)]a^\dagger + U_\omega b(\omega)d^\dagger + \Hc
\end{align}
\end{subequations}
Next, we formulate the equations of motion for the four fields,
\begin{subequations}
\begin{align}
    \dot{a} &= -i\omega_a a-i\lambda e^{ i\phi}d + \int d\omega\,[V_\omega b(\omega)+W_\omega c(\omega)],\label{eq:AppCavity}\\
    \dot{d} &= -i\omega_d d-i\lambda e^{-i\phi}a + \int d\omega\,U_\omega b(\omega),\label{eq:AppFano}\\
    \dot{b}(\omega) &= -i\omega b(\omega) - V_\omega a - U_\omega d,\\
    \dot{c}(\omega) &= -i\omega c(\omega) - W_\omega a,
\end{align}
\end{subequations}
and formally integrate the equations for the two continua,
\begin{subequations}\label{eq:Continua}
\begin{align}
    b(\omega) &= e^{-i\omega(t-t_0)}b_0(\omega) -\int_{t_0}^t d\tau\,e^{-i\omega(t-\tau)}(V_\omega a+U_\omega d),\\
    c(\omega) &= e^{-i\omega(t-t_0)}c_0(\omega) -\int_{t_0}^t d\tau\,e^{-i\omega(t-\tau)}W_\omega a,
\end{align}
\end{subequations}
where we assume the initial conditions $b_0(\omega)$, $c_0(\omega)$ at time $t_0\to -\infty$.

Now, we plug the results~\eqref{eq:Continua} back to the equations of motion for the cavity field and mirror mode.
We assume that the response of the mirrors is sufficiently flat for the frequencies of interest so that we can assume the tunneling rates $U,V,W$ to be frequency independent.
We can then evaluate the integrals over frequency in Eqs.~\eqref{eq:AppCavity}, \eqref{eq:AppFano} (each of which yields a delta function) and obtain
\begin{subequations}\label{eq:SM_EOM}
\begin{align}
    \dot{a} &= -(i\omega_a +\kappa_L+\kappa_R)a -(i\lambda e^{i\phi}+\sqrt{\kappa_L\gamma})d + \sqrt{2\kappa_L}b_\inpt + \sqrt{2\kappa_R}c_\inpt\label{eq:AppAFUll}\\
    \dot{d} &= -(i\omega_d+\gamma)d -(i\lambda e^{-i\phi}+\sqrt{\kappa_L\gamma})a + \sqrt{2\gamma}b_\inpt;
\end{align}
\end{subequations}
here, we introduced the decay rates $\kappa_L = \pi V^2$, $\kappa_R =\pi W^2$, $\gamma =\pi U^2$ and the input field
\begin{equation}
    b_\inpt = \frac{1}{\sqrt{2\pi}}\int d\omega\, e^{-i\omega(t-t_0)}b_0(\omega);
\end{equation}
an analogous definition holds also for $c_\inpt$.

Alternatively, the equations of motion for the two continua can be solved using the final conditions $b_1(\omega)$, $c_1(\omega)$ at time $t_1\to \infty$,
\begin{subequations}
\begin{align}
    b(t,\omega) &= e^{-i\omega(t-t_1)}b_1(\omega) +\int_{t}^{t_1} d\tau\,e^{-i\omega(t-\tau)}(V_\omega a+U_\omega d),\\
    c(t,\omega) &= e^{-i\omega(t-t_1)}c_1(\omega) +\int_{t}^{t_1} d\tau\,e^{-i\omega(t-\tau)}W_\omega a.
\end{align}
\end{subequations}
Introducing the output field
\begin{equation}
    b_\out = \frac{1}{\sqrt{2\pi}}\int d\omega\, e^{-i\omega(t-t_1)}b_1(\omega)
\end{equation}
and expressing
\begin{align}
\begin{split}
    \frac{1}{\sqrt{2\pi}}\int d\omega\,b(\omega) &= b_\inpt - \sqrt{\frac{\kappa_L}{2}}a - \sqrt{\frac{2}{\gamma}}d \\
    &= b_\out +\sqrt{\frac{\kappa_L}{2}}a + \sqrt{\frac{2}{\gamma}}d,
\end{split}
\end{align}
we obtain the input--output relations
\begin{subequations}\label{eq:InOut}
\begin{align}
    b_\out &= b_\inpt - \sqrt{2\kappa_L}a - \sqrt{2\gamma}d,\label{eq:AppInOutB}\\
    c_\out &= c_\inpt - \sqrt{2\kappa_R}a.\label{eq:AppInOutc}
\end{align}
\end{subequations}
The input--output relation~\eqref{eq:AppInOutc} can be derived in full analogy to Eq.~\eqref{eq:AppInOutB}.
Note that, in the equations above, the description of a standard two-sided cavity can be recovered by putting $\lambda = U = \gamma = 0$.

Non-Markovian dynamics of the cavity field can now be obtained by solving the equation for the mirror mode formally,
\begin{equation}
    d(t)=\int_{-\infty}^{t}d\tau\, e^{-(i\omega_d+\gamma)(t-\tau)}[\sqrt{2\gamma}b_\inpt(\tau)- \mathcal{G}_-a(\tau)],
\end{equation}
and plugging this solution into the equation of motion for the cavity field, Eq.~\eqref{eq:AppAFUll}, and the input--output relation~\eqref{eq:AppInOutB}.
We obtain
\begin{subequations}
\begin{align}
    \dot{a} &= -i\omega_a a -\kappa_R a+\sqrt{2\kappa_R}c_\inpt -\int_{-\infty}^t d\tau\,[\kappa_\eff(t-\tau)a(\tau)+ \kappa_\inpt(t-\tau)b_\inpt(\tau)], \\
    b_\out &= \int_{-\infty}^t d\tau\,[\kappa_\out(t-\tau)b_\inpt(\tau) - \kappa_\inpt'(t-\tau)a(\tau)];
\end{align}
\end{subequations}
the coefficients $\kappa_\eff$, $\kappa_\inpt$, $\kappa_\inpt'$, $\kappa_\out$ are given by
\begin{subequations}\label{eq:Coefficients}
\begin{align}
    \kappa_\eff(t) &= 2\kappa_L\delta(t)-\mathcal{G}_+\mathcal{G}_-e^{-(i\omega_d+\gamma)t}\\
    \kappa_\inpt(t) &= 2\sqrt{2\kappa_L}\delta(t)-\mathcal{G}_+\sqrt{2\gamma}e^{-(i\omega_d+\gamma)t},\\
    \kappa_\inpt'(t) &= 2\sqrt{2\kappa_L}\delta(t)-\mathcal{G}_-\sqrt{2\gamma}e^{-(i\omega_d+\gamma)t},\\
    \kappa_\out(t) &= 2\delta(t)-2\gamma e^{-(i\omega_d+\gamma)t}.
\end{align}
\end{subequations}

For photonic crystal membrane cavities, we derive below
\begin{subequations}
\begin{align}
    \mathcal{G}_+ &= \mathcal{G}_- = \mathcal{G} = \sqrt{\kappa_L\gamma} -is \sqrt{\kappa_0\gamma},\\
    \omega_a &= \omega_d-2s\sqrt{\kappa_0\kappa_L}.
\end{align}
\end{subequations}
We can use these expressions to obtain the effective time-dependent decay rates
\begin{subequations}\label{eq:NMarkRates}
\begin{align}
    \kappa_\eff(t) &= 2\kappa_L\delta(t)-\mathcal{G}^2K(t),\\
    \kappa_\inpt(t) = \kappa_\inpt'(t) &= 2\sqrt{2\kappa_L}\delta(t)-\mathcal{G}\sqrt{2\gamma}K(t),\\
    \kappa_\out(t) &= 2\delta(t)-2\gamma K(t),
\end{align}
\end{subequations}
where we introduced the kernel function $K(t) = \exp[-(i\omega_d+\gamma)t]$.
In the limit $\gamma\to\infty$, we have $\gamma e^{-\gamma t}\to 2\delta(t)$ so the rates defined in Eqs.~\eqref{eq:NMarkRates} become proportional to the delta function.
We thus recover the expected Markovian dynamics
\begin{subequations}
\begin{align}
    \dot{a} &= -(i\omega_d+\kappa_R+\kappa_0) a + i\sqrt{2\kappa_0}b_\inpt + \sqrt{2\kappa_R} c_\inpt,\\
    b_\out &= -b_\inpt +i\sqrt{2\kappa_0}a;
\end{align}
\end{subequations}
the resonance frequency of the cavity is shifted from $\omega_a$ to $\omega_d$ in agreement with the transfer matrix calculation (see the next section).

\section{Comparison to transfer matrix description}

In this section, we show how the parameters of the coupled-mode model can be related to experimentally accessible quantities.
We start by evaluating the transmission function from the coupled-mode theory.
First, we solve the equations of motion~\eqref{eq:SM_EOM} in the frequency domain; this yields the result
\begin{subequations}\label{eq:FTSolution}
\begin{align}
    a(\Delta) &= \frac{[(\gamma+i\Delta)\sqrt{2\kappa_L}-\mathcal{G}_+\sqrt{2\gamma}]b_\inpt+(\gamma+i\Delta)\sqrt{2\kappa_R}c_\inpt}{(\kappa+i\Delta+i\delta)(\gamma+i\Delta)-\mathcal{G}_+\mathcal{G}_-},\\
    d(\Delta) &= \frac{[(\kappa+i\Delta+i\delta)\sqrt{2\gamma}-\mathcal{G}_-\sqrt{2\kappa_L}]b_\inpt-\mathcal{G}_-\sqrt{2\kappa_R}c_\inpt}{(\kappa+i\Delta+i\delta)(\gamma+i\Delta)-\mathcal{G}_+\mathcal{G}_-}
\end{align}
\end{subequations}
with the detunings $\Delta = \omega_d-\omega$, $\delta = \omega_a-\omega_d$.
Next, we plug Eqs. \eqref{eq:FTSolution} into the input--output relations~\eqref{eq:InOut}.
We thus obtain an expression that describes the transmission from the input field $b_\inpt$ to the output $c_\out$,
\begin{equation}\label{eq:Transmission}
    t(\Delta) = \frac{2i\sqrt{\kappa_R}[e^{i\phi}\lambda\sqrt{\gamma}-\Delta\sqrt{\kappa_L}]}{\lambda^2+\kappa_R\gamma-\Delta(\Delta+\delta)+i[\gamma(\Delta+\delta)+\kappa\Delta-2\lambda\sqrt{\kappa_L\gamma}\cos\phi]}.
\end{equation}

In the transfer matrix formalism~\cite{Naesby2018}, each mirror can be characterized using its polarizability $\zeta_i$, $i = R,L$.
For the right mirror, the amplitude transmission and reflection coefficients are given by $t_R = 1/(1-i\zeta_R)$, $r_R = i\zeta_R/(1-i\zeta_R)$ and the associated transfer matrix by
\begin{equation}\label{eq:TMRight}
    M_R = \left(\begin{array}{cc}
        1+i\zeta_R & i\zeta_R \\ -i\zeta_R & 1-i\zeta_R
    \end{array}\right).
\end{equation}
An analogous expression holds also for the left mirror with general frequency-dependent polarizability $\zeta_L(\Delta)$.
Finally, the transfer matrix for free propagation between the two mirrors is given by $M_f = \diag(e^{i\theta},e^{-i\theta})$,
where $\theta = \omega/2\Gamma$ and $\Gamma = c/2L$ is the free spectral range.

The transfer matrix of the cavity is obtained by multiplying all three matrices,
\begin{align}\label{eq:TransferMatrix}
\begin{split}
    M(\Delta) &= M_R M_f M_L \\
        &= \left(\begin{array}{cc}
            (1+i\zeta_R)[1+i\zeta_L(\Delta)]e^{i\theta}+\zeta_R\zeta_L(\Delta)e^{-i\theta} & i[1-i\zeta_L(\Delta)]\zeta_Re^{-i\theta}+i\zeta_L(\Delta)(1+i\zeta_R)e^{i\theta} \\
            -i[1+i\zeta_L(\Delta)]\zeta_Re^{i\theta}-i\zeta_L(\Delta)(1-i\zeta_R)e^{-i\theta} & (1-i\zeta_R)[1-i\zeta_L(\Delta)]e^{-i\theta}+\zeta_R\zeta_L(\Delta)e^{i\theta}
        \end{array}\right).
\end{split}
\end{align}
The cavity transmission coefficient can now be obtained as $\tilde{t}(\Delta) = 1/m_{22}(\Delta)$ which yields
\begin{equation}\label{eq:TransTM}
    \tilde{t}(\Delta)
    = \frac{1}{(1-i\zeta_R)[1-i\zeta_L(\Delta)]e^{-i\theta}+\zeta_R\zeta_L(\Delta)e^{i\theta}}.
\end{equation}
In going from the general transfer matrix, Eq.~\eqref{eq:TransferMatrix}, to Eq.~\eqref{eq:TransTM}, we expressed the exponentials $e^{\pm i\theta}$ in terms of the detuning $\Delta = \omega_d-\omega$ instead of the frequency $\omega$.
This step is possible when one notes that the cavity field fulfills the resonance condition
\begin{equation}
    2\frac{\omega_dL}{c} = \arctan \frac{1}{\zeta_R}+\arctan \frac{1}{\zeta_0}.
\end{equation}

For the example of photonic crystal membranes, we have the polarizability~\cite{Naesby2018}
\begin{equation}\label{eq:zetaL}
    \zeta_L(\Delta) = \zeta_0\frac{\gamma-2s'\Delta/\zeta_0}{\gamma+2s'\zeta_0\Delta}.
\end{equation}
Here, the parameter $s' = \pm 1$ determines the orientation of the Fano resonance in the mirror's reflectivity.
We also suppose that the resonance frequencies of the cavity mode and the polarizability $\zeta_L(\Delta)$ coincide.
In this regime, the transmission is maximized for $\Delta = 0$ and reaches unity for $\zeta_0 = \zeta_R$.
Note that this condition is different from the coupled-mode description where we assume a general detuning $\delta$ between the two.
This discrepancy stems from the different polarizabilities used to determine the cavity resonance frequency:
For the coupled-mode model, the relevant polarizability of the left mirror is $1/\zeta_0$ whereas it is $\zeta_0$ for the transfer matrix calculation.
We also identified the spectral width of the polarizability~\eqref{eq:zetaL} with the linewidth of the mirror mode $d$;
this result follows from coupled-mode analysis of the membrane in the absence of a second mirror~\cite{Fan2003}.
We start the identification by making several observations:

(i)
When $\Delta = -s'\gamma/2\zeta_0$ in the transfer matrix calculation, the polarizability diverges, $\zeta_L(\Delta)\to\infty$;
the reflectivity reaches unity and there is no transmission.
To achieve the same effect in the coupled-mode theory, we must have $\phi = k\pi$, $k\in\mathbb{Z}$ as only then can the numerator in Eq.~\eqref{eq:Transmission} vanish.
The coupling rate
\begin{equation}\label{eq:FanoZero}
    \lambda = -\frac{s}{2\zeta_0}\sqrt{\kappa_L\gamma}
\end{equation}
with $s = e^{i\phi} = \pm 1$ then ensures zero transmission for the same detuning.

(ii)
When the width of the reflectivity is larger than the free spectral range, $\gamma\gg\Gamma$, the polarizability $\zeta_L(\Delta)\simeq\zeta_0$ for all frequencies of interest.
We then recover the Lorentzian transmission
\begin{equation}\label{eq:BadFanoTM}
    \tilde{t}(\Delta) = \frac{\tilde\kappa e^{i\varphi}}{\tilde\kappa+i\Delta}
\end{equation}
with the linewidth $\tilde\kappa = \Gamma/\zeta_0^2$ and a polarizability-dependent phase $\varphi$; for $\zeta_0\gg 1$, one finds $e^{i\varphi}\simeq-i$.
The transmission derived from the coupled mode theory in the limit $\gamma\to\infty$ becomes
\begin{equation}
    t(\Delta) = \frac{-i\sqrt{\kappa_R\kappa_L}/\zeta_0}{\kappa_R+\kappa_L/4\zeta_0^2 + i(\Delta+\delta+s\kappa_L/\zeta_0)}.
\end{equation}
From the real part of the denominator, we must have $\tilde\kappa = \kappa_R+\kappa_L/4\zeta_0^2$;
moreover, the cavity decays through both mirrors equally if $\kappa_R = \Gamma/2\zeta_0^2$ and $\kappa_L = 2\Gamma$.
With these values, the numerator becomes $-2i\kappa_R$.
Finally, from the imaginary part of the denominator, we get $\delta = -s\kappa_L/\zeta_0 = -2s\sqrt{\kappa_R\kappa_L}$.

(iii)
For very narrow polarizability, $\gamma\ll\Gamma$, the intensity transmission can be approximated as~\cite{Naesby2018}
\begin{equation}\label{eq:GoodFanoTM}
    |\tilde{t}(\Delta)|^2 \simeq \left(1 + (1+\zeta_0^2)^2\frac{4\Delta^2}{(\gamma+2s'\zeta_0\Delta)^2}\right)^{-1};
\end{equation}
it has an asymmetric profile with linewidth $\gamma/2\zeta_0^2$.
The coupled mode approach in the same limit (i.e., with $\gamma$ being the smallest rate in the system) yields
\begin{equation}\label{eq:GoodFano}
    |t(\Delta)|^2 \simeq \left(1 + \frac{\kappa_L^2\Delta^2}{4(\kappa_R\gamma+s\Delta\sqrt{\kappa_R\kappa_L})^2}\right)^{-1}.
\end{equation}
The transmission spectra in Eqs.~\eqref{eq:GoodFanoTM} and~\eqref{eq:GoodFano} are approximately equal provided $\zeta_0\gg 1$ and $s = s'$;
the former condition is necessary in order to have a well-defined cavity mode so it does not restrict the validity of our approach.

We derived the above results under the assumption that the resonant polarizability of the left mirror is equal to the polarizability of the right mirror, $\zeta_0 = \zeta_R$.
Our results can, however, be easily generalized to situations where this is not the case:
The decay rate associated with the right mirror is given by $\kappa_R = \Gamma/2\zeta_R^2$ as one would expect and $\kappa_L = 2\Gamma$ remains unchanged.
The coupling of the mirror and cavity modes $\lambda$, as well as the detuning of the cavity $\delta$, can be expressed in terms of the left-mirror polarizability $\zeta_0$ as
\begin{equation}
    \lambda = -s\sqrt{\kappa_0\gamma},\qquad\delta = -2s\sqrt{\kappa_0\kappa_L}.
\end{equation}
With these modifications, one can also recover the dynamics of a one-sided cavity, for which $\zeta_R\to\infty$ (or $\kappa_R\to 0$).

\begin{figure}
    \centering
    \includegraphics[width=0.5\linewidth]{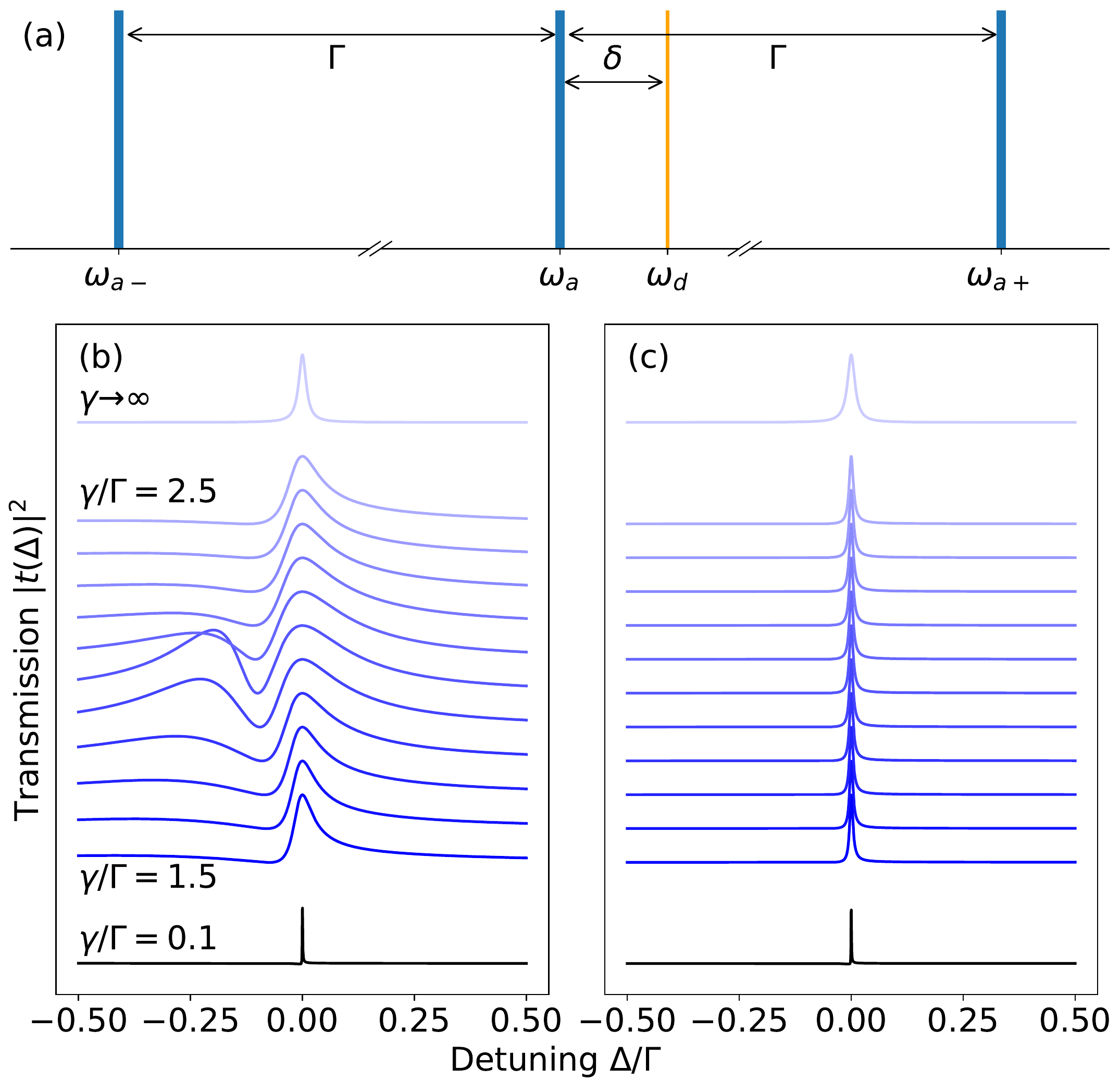}
    \caption{\label{fig:Intermediate}
        (a) Modes involved in the coupled-mode model.
        Apart from the cavity mode $a$ at frequency $\omega_a$, we show also the two neighboring cavity modes at frequencies $\omega_{a\pm} = \omega_a \pm\Gamma$ (thick blue lines);
        the mirror's internal mode at frequency $\omega_d$ is depicted as the thin orange line.
        (b,c) Transmission spectra obtained from (b) the transfer matrix calculation and (c) coupled-mode model for various widths of the mirror mode.
        We use the parameters $\zeta_0 = \zeta_R = 10$, $s = 1$;
        the offset in the $y$ direction serves to improve readability of the data.
    }
\end{figure}

We plot the results of the transfer matrix calculation in Fig.~\ref{fig:Intermediate}(b) for various values of the mirror mode linewidth.
In the limit $\gamma\ll\Gamma$, we obtain the asymmetric lineshape of width $\kappa_\eff = \gamma/2\zeta_0^2$ and, for $\gamma\to\infty$, we obtain the Lorentzian profile of width $\Gamma/\zeta_0^2$; we discuss these results in the main text.
In the intermediate regime, $\gamma\sim\Gamma$, the transmission profile becomes much broader which cannot be replicated in the coupled-mode model with a single cavity mode [see Fig.~\ref{fig:Intermediate}(c)].
This discrepancy stems from coupling of the mirror mode to the neighboring cavity modes at frequencies $\omega_{a\pm} = \omega_a\pm\Gamma$ [as shown in Fig.~\ref{fig:Intermediate}(a)].
When the mirror mode linewidth is comparable to the free spectral range, the mirror mode can mediate interactions between different cavity modes that result in an additional decay of the cavity mode $a$, leading to a broadened transmission spectrum.
This effect can be included in the coupled-mode model by adding the modes $a_\pm$ and deriving their coupling to the mirror mode.
This regime is, however, not particularly interesting for applications.

\section{Optomechanical sideband cooling}

To evaluate the spectral density of the Langevin force $F = g(a+a^\dagger)$ responsible for cooling, $S_F(\omega)$, we solve the equations of motion for the cavity and mirror modes perturbatively, in that we neglect the backaction of the mechanical oscillator on the two modes.
We thus have the Langevin equations
\begin{subequations}
\begin{align}
    \dot{a} &= -(i\Delta_a+\kappa)a -\mathcal{G} d + \sqrt{2\kappa_L}b_\inpt + \sqrt{2\kappa_R}c_\inpt,\\
    \dot{d} &= -(i\Delta_d+\gamma)d -\mathcal{G} a + \sqrt{2\gamma}b_\inpt;
\end{align}
\end{subequations}
the solution for the cavity mode in the frequency space can be written as
\begin{equation}
    a(\omega) = \tilde{\chi}_a(\omega)[(\sqrt{2\kappa_L}-\mathcal{G}\chi_d(\omega)\sqrt{2\gamma})b_\inpt(\omega) + \sqrt{2\kappa_R}c_\inpt(\omega)].
\end{equation}
Here, we introduced the bare susceptibilities $\chi_a^{-1}(\omega) = \kappa-i(\omega-\Delta_a)$, $\chi_d^{-1}(\omega) = \gamma-i(\omega-\Delta_d)$ and the dressed susceptibility $\tilde{\chi}_a^{-1}(\omega) = \chi_a^{-1}(\omega) - \mathcal{G}^2\chi_d(\omega)$.
With this solution and the correlation functions for the input fields $b_\inpt$, $c_\inpt$, the spectral density becomes
\begin{align}
    S_F(\omega) &= 2g^2|\tilde{\chi}_a(\omega)|^2 [\kappa_R + |\sqrt{\kappa_L}-\mathcal{G}\chi_d(\omega)\sqrt{\gamma}|^2]\nonumber \\
    &= \frac{2g^2\kappa_R[\gamma^2+(\omega-\Delta_d)^2]+2g^2[\gamma\sqrt{\kappa_0}-s\sqrt{\kappa_L}(\omega-\Delta_d)]^2}{[(\kappa+\gamma)(\omega-\Delta_d)]^2 + [\gamma(\kappa_0+\kappa_R)-(\omega-\Delta_d)(\omega-\Delta_d+2s\sqrt{\kappa_0\kappa_L})]^2}.
\end{align}

To find the final mechanical occupation numerically, we solve the Lyapunov equation for the covariance matrix of the system.
As we describe in detail elsewhere~\cite{Cernotik2018}, one proceeds by collecting the quadrature operators in the vector $r = (X_a,Y_a,X_d,Y_d,q,p)^T$,
where $X_a = (a+a^\dagger)/\sqrt{2}$, $Y_a = -i(a-a^\dagger)/\sqrt{2}$ and similar for $d$.
Since the dynamics of the system are linear, the time evolution can be fully described by an equation of motion for the covariance matrix,
\begin{equation}
    \dot{V} = AV+VA^T+N,
\end{equation}
where the matrices $A,N$ depend on the form of the Hamiltonian and the Lindblad terms;
for our system, we have
\begin{subequations}
\begin{align}
    A &= \left(\begin{array}{cccccc}
        -\kappa & \Delta_a & -\sqrt{\kappa_L\gamma} & -s\sqrt{\kappa_0\gamma} & 0 & 0\\
        -\Delta_a & -\kappa & s\sqrt{\kappa_0\gamma} & -\sqrt{\kappa_L\gamma} & -2g & 0\\
        -\sqrt{\kappa_L\gamma} & -s\sqrt{\kappa_0\gamma} & -\gamma & \Delta_d & 0 & 0\\
        s\sqrt{\kappa_0\gamma} & -\sqrt{\kappa_L\gamma} & -\Delta_d & -\gamma & 0 & 0\\
        0 & 0 & 0 & 0 & -\gamma_m & \omega_m \\
        -2g & 0 & 0 & 0 & -\omega_m & -\gamma_m
    \end{array}\right),\\
    N &= \left(\begin{array}{cccccc}
        2\kappa & 0 & 2\sqrt{\kappa_L\gamma} & 0 & 0 & 0\\
        0 & 2\kappa & 0 & 2\sqrt{\kappa_L\gamma} & 0 & 0\\
        2\sqrt{\kappa_L\gamma} & 0 & 2\gamma & 0 & 0 & 0\\
        0 & 2\sqrt{\kappa_L\gamma} & 0 & 2\gamma & 0 & 0\\
        0 & 0 & 0 & 0 & 2\gamma_m(2\nbar+1) & 0\\
        0 & 0 & 0 & 0 & 0 & 2\gamma_m(2\nbar+1)
    \end{array}\right);
\end{align}
\end{subequations}
here, we introduced the initial mechanical occupation at a temperature $T$, $\nbar \simeq k_BT/\hbar\omega_m$.
We find the steady state covariance matrix $\bar{V}$ (which is possible provided all eigenvalues of $A$ have negative real parts) and determine the final mechanical occupation from the variance of mechanical position and momentum, $n_f = (\bar{V}_{55}+\bar{V}_{66}-2)/4$.

\end{widetext}

\end{document}